\documentclass[12pt]{article}  %\documentstyle[12pt]{article}
\usepackage{graphicx}
\usepackage{amsmath}
\usepackage[margin=0.9in]{geometry}	
\newcommand{\B}{Ba\v zant}  %\newcommand{\BB}{Ba\v zant }
\newcommand{\vv}{\vspace*{1.5ex}}

                            \newcommand{\no}{\noindent}
 \newcommand{\bc}{\begin{center}}
 \newcommand{\ec}{\end{center}}
                   \newcommand{\bfr}{\begin{flushright}}
                   \newcommand{\efr}{\end{flushright}}
   \newcommand{\ii}{\item}
     \newcommand{\be}{\begin{enumerate}}
     \newcommand{\ee}{\end{enumerate}}
        \newcommand{\bi}{\begin{itemize}}
        \newcommand{\ei}{\end{itemize}}
            \newcommand{\bd}{\begin{description}}
            \newcommand{\ed}{\end{description}}
                \newcommand{\beq}{\begin{equation}}
                \newcommand{\eeq}{\end{equation}}
                  \newcommand{\bea}{\begin{eqnarray}}
                  \newcommand{\eea}{\end{eqnarray}}

      \newcommand{\bfi}{\begin{figure}}
      \newcommand{\efi}{\end{figure}}
\newcommand{\bay}{\begin{array}{l}}
\newcommand{\eay}{\end{array}}
            
    \newcommand{\pa}{\partial}
    
    \newcommand{\Del}{\Delta}
    
    \newcommand{\la}{\lambda}
    
    \newcommand{\al}{\alpha}
    \newcommand{\sig}{\sigma}
    \newcommand{\eps}{\epsilon}
    \newcommand{\ga}{\gamma}

 \newcommand{\ssig}{\mbox{\boldmath{$\sigma$}}}

\newcommand{\CC}{\mbox{\boldmath $C$}}
\begin{document}   
   \thispagestyle{empty}
        \hspace*{1mm}  \vspace*{-0mm}
\noindent {\footnotesize {{\em
\hfill      Posted on ArXiv}} %, submitted to JEM-ASCE} }}
    \vskip 1.5in
\begin{center}
 {\Large {\sf  Critique of Critical Shear Crack Theory (CSCT) for {\em fib} Model Code Articles on Shear Strength and Size Effect of RC Beams  \\[7mm]
}}

{\large {\sc Abdullah A. D\" onmez and Zden\v ek P. Ba\v zant}}
\\[1.5in]

 {\Large{\sf SEGIM Report No. 18-10/788c}}\\[2.1in]
%Here 33788 refers to my expired DoE Grant on ASR. ZPB
   %NOTE: 778 is the abbreviated project number; 09 stands for the
   %month, x stands for the first letter of the title, or other
   %if there is a confusion.

Center for Structural Engineering of Geological and Infrastructure Materials (SEGIM)
\\ Department of Civil and Environmental Engineering
%McCormick School of Engineering and Applied Science
%\\ Departments of Civil Engineering and Materials Science??
\\ Northwestern University
\\ Evanston, Illinois 60208, USA
\\[1in]  {\bf October 26, 2018} %\\ Revised ... 2003
\end{center}
   %NOTE: Adjust spacing to fit one page, if needed

\clearpage   \pagestyle{plain} \setcounter{page}{1}
   %NOTE: This is necessary only for insertion into Latex paper

\begin{center}
 {\Large {\sf  Critique of Critical Shear Crack Theory (CSCT) for {\em fib} Model Code Articles on Shear Strength and Size Effect of RC Beams  \\[7mm]
 }}
 %and Energy Based Comparisons  }}
                % Is CSCT Realistic Basis for Size Effect on Shear Strength of RC Beams?
            % for Size Effect on Concrete Shear Strength
            % and Comparison to Fracture Mechanics
     %Problems in Critical Shear Crack Width Hypothesis (CSCT Theory) for Concrete %Design Code and Comparison to Quasibrittle Fracture Mechanics
 {\large {\bf
    Abdullah D\"onmez\footnote{Postdoctoral Research Associate, Istanbul Technical University, Istanbul, Turkey and Visiting Postdoc at Northwestern University}
           and
    Zden\v ek P. Ba\v zant}\footnote{
McCormick Institute Professor and W.P. Murphy Professor of Civil
and Mechanical Engineering and Materials Science, Northwestern University, 2145 Sheridan Road, CEE/A135, Evanston, Illinois 60208; corresponding author,
z-bazant@northwestern.edu.} }
    %\\(first draft by ZPB, 8/1/18)
\end{center} \vskip 5mm   \baselineskip 13.8pt

\noindent {\bf Abstract:}\, {\sf The size effect of Muttoni et al.’s CSCT is shown to  be very close and asymptotically identical to the energetic size effect law (SEL) extensively verified experimentally and theoretically  (due to appear in the 2019 ACI Code for both beam shear and punching). However, the CSCT derivation and calculation procedure obfuscates the mechanics of failure. It is shown to rest on six scientifically untenable hypotheses, which makes CSCT untrustworthy outside the testing range. It would have to be taught to students as a dogma. These conclusions are supported by experimentally calibrated finite element simulations of crack path and width, stress distributions and localizations during failure, and strain energy release. The simulations also show the CSCT to be incompatible with the `strut-and-tie' model modernized to include the size effect in the compression strut. Finally, further deficiencies are pointed out for the Modified Compression Field Theory (MCFT), currently embedded in the Model Code.
}

\subsection*{\large Nature and Evolution of Size Effect Formulae in Design Codes}    %Introduction}

Half a century ago, experiments in Stuttgart \cite{LeoWal62,LeoWaLDil1964,Bhal}, Toronto \cite{Kan66, Kan67}, and Tokyo \cite{IguShi-1985} established the existence of a strong size effect in shear failure of reinforced concrete (RC) beams. Weibull's \cite{Wei39} statistical power-law size effect was already well known at that time, but had also been known to apply only to structures in which formation of a small crack (or fractured representative volume element of material) within any one of many possible places in the structure volume generates a dynamic crack propagation and causes immediate failure. This is obviously not the case for shear failure of RC beams, which tolerate extensive cracking and a long stable crack before the ultimate load.

Based on energy release arguments adapted to quasibrittle fracture mechanics, a new energetic size effect law (SEL), applicable to failures occurring after stable growth of a long crack, as typical of shear failure of RC beams, was formulated in 1984 \cite{Baz84}. Immediately \cite{BazKim84}, the SEL was proposed to ACI for shear design of RC beams (as well as of prestressed beams \cite{BazCao86}). Subsequently it was shown to apply to many types of failure in all quasibrittle materials \cite{BazPla98, Baz05}, which do not follow classical fracture mechanics. Aside from concrete, they also include tough ceramics, fiber composites, rocks and stiff soils, sea ice, wood, stiff foams, etc. The SEL captures the transition from nearly ductile behavior in small concrete structures to nearly brittle behavior in large ones. The reason for this ductile-brittle transition of structural response is the material heterogeneity, which causes the fracture process zone (FPZ) to be long (cca 0.5 m in concrete, vs. micrometers in metals), and non-negligible compared to the cross section size.

The size effect theory based on quasibrittle fracture mechanics had to wait three decades to win broad acceptance in the engineering community. This long delay was doubtless caused by various controversies generated by competing explanations of size effect---e.g., the fractal nature of crack surfaces or of microstructure of concrete, the role of boundary layer, and the effects of initial crack spacing, of crack width, or of material randomness of various kinds.

Lacking, in the 1980s, a generally accepted theory of quasibrittle failures, the Japan Society of Civil Engineers (JSCE) and Commit\' e Europ\' een de b\' eton (CEB), apparently thinking that `better something than nothing', introduced three decades ago
 %9/6 Mr. Donmez, better give references and years if your could find them, ZPB %AD-I couldnt find the earlier versions.So I used the current ones
into their design specifications purely empirical size effect equations, very different from each other \cite{jsce, MC2010}. Meanwhile, ACI committees, cognizant of the enormous staying power of code specifications adopted by democratic voting of large committees, and thinking cautiously that `better nothing than something controversial', spent three decades in lively polemics until eventually deciding to adopt the SEL for the the beam shear and slab punching specifications (which are now approved to appear in the 2019 version of ACI code, Standard 318).

Meanwhile, {\em fib} (f\' ed\' eration internationale de b\' eton), the successor to CEB, made a change in its Model Code 2010 \cite{MC2010}. It adopted the size effect equations based on the so-called Modified Field Compression Theory" (MCFT) \cite{CollinsVechio, MC-Background}. The MCFT uses elementary, supposedly logical, arguments, in which the critical crack width, $w$, and a certain strain, $\eps_x$, estimated from the classical elastic theory of beam bending, are imagined to be the failure indicators.
						
Model Code 2010 \cite{MC2010} specifies in its equations 7.3-19 and 7.3-21 two approximations for beam shear capacity without stirrups---Level I and Level II. This paper will focus on Level II approximation. Level I, which is treated (according to Eq. 7.3-21 in \cite{MC2010}) as an approximation to Level II, will be only briefly examined in the Appendix.

Currently, the so-called Critical Shear Crack Theory (CSCT) \cite{MutRui08, RuiMut18, RuiMut15}, extending a concept advanced in 1991 \cite{csct}, is being proposed as an improvement of Model Code 2010 for beam (or one-way) shear.        %and punching (two-way}shear.
The Swiss Code \cite{SwissCode} has already adopted the CSCT for both the beam shear and punching (two-way) shear,  %9/11 is the line below true?
while Model Code 2010 has done so for the latter. The objective of this article is to call attention to the errors CSCT.

%, and particularly to show that the CSCT equations are unrealistic and rest on illogical arguments conflicting with fracture mechanics.
  %AD ends - modified %9/6

\subsection*{\large Hypotheses Underlying the CSCT Theory}

Careful examination shows that six hypotheses are implicit to CSCT.   %The
%CSCT theory may be clarified by dissecting it into a set of six hypotheses.
The average (or nominal) shear strength (or ultimate stress) in the cross section is denoted as $v_u = V_R /bd$ where $V_R$ is the resistant shear force provided by concrete; $d$ = depth of the cross section from the compressed face to the centroid of reinforcement; $b$ = width.

{\bf Hypothesis 1.} The shear force, $V_R$, carried by concrete at maximum load is (in CSCT as well as MCFT) assumed to be controlled by a  characteristic crack width $w$ of the dominant crack leading to failure.

{\bf Hypothesis 2.} To express the size effect, it is assumed that (Eq. (1) in \cite{RuiMut18}):
 \beq \label{1}
  \frac{v_u}{\sqrt{f_c}} = \frac{\al_1}{1 + \al_2 (w / d_{dg})},~~~~
  \al_1 = 1/3
  ~~~~\mbox{(in MPa, mm)}
 \eeq
where $f_c$ is the mean compression strength of concrete (both $f_c$ and $\sqrt{f_c}$ is considered to be in MPa); $d_{dg}$ is called the equivalent surface roughness and is calculated as $d_{dg} =$ min($d_g$ + 16, 40 mm).

{\bf Hypothesis 3:} The width of the dominant diagonal crack is assumed to be proportional to the reference strain, $\eps$, i.e.,
 \beq \label{2}
  w\ =\ \al_3\, \eps\, d,~~~~\al_2 \al_3 = 120~\mbox{mm}
 \eeq
where $\al_2, \al_3$ are empirical coefficients and $\eps$ is called the reference strain, defined as the normal strain in the longitudinal direction, $x$, at a certain characteristic location, crossing the dominant crack (the on-line equation above Eq. 1 in \cite{RuiMut18}).

{\bf Hypothesis 4:} The reference strain, $\eps$, is assumed to be the longitudinal normal strain at distance $d/2$ from the concentrated load $P = V_R$ toward the support, and at depth $0.6 d$ from the top face of beam (Eq. 2  in \cite{RuiMut18}).

{\bf Hypothesis 5:} The reference strain, $\eps$, assumed to control crack width, is calculated according to the linear elastic beam theory based on Bernoulli-Navier hypothesis of plane cross sections remaining plane (Eqs. 2 and 3 in \cite{RuiMut18}), i.e.,
  \beq \label{3}
   \eps = \frac M {b d \rho E_s (d - c/3)}\ \frac{\al_4 d - c}{d - c},~~~~
   M = V_R (a - \al_5 d), ~~~\al_4 = 0.6,~~\al_5 = 1/2
  \eeq
and, according to the no-tension hypothesis of elastic flexure of concrete beams with one-sided reinforcement (\cite{RuiMut18} Eq. (4)),
  \beq \label{4}
   c = \al_6 d,~~~\al_6 =
   \frac{\rho E_s}{E_c}\left( \sqrt{1 + \frac{2 E_c}{\rho E_s}}\ - 1\, \right)
  \eeq
where $E_c, E_s$ = elastic moduli of concrete and of steel reinforcement, $\rho$ = reinforcement ratio, $c$ = distance of the neutral axis from the top face of beam (or the length of the triangular profile of compression stress), $a$ = shear span = distance between the concentrated load and the closest end support.

{\bf Hypothesis 6.} The foregoing equations based on linear elastic beam bending theory are assumed to be applicable at maximum (or ultimate) load of beam, i.e., at incipient shear failure under controlled load.

\subsection*{\large Beam Shear Strength According to CSCT Ensuing from the Hypotheses, and Comparison with ACI-446}

In dimensionless form, $\eps$ in Eq. (\ref{2}) (Hypotheses 3, 5, 6) can be written as
  \beq  \label{5}
  \eps =  \ga v_u,~~~~
  \ga = \frac {a- \al_5 d} {\rho E_s (d - c/3)}\ \frac{\al_4 d - c}{d - c}
  \eeq
Substituting $\eps$ from Eq. (\ref{2}) into Eq. (\ref{1}) (Hypotheses 1, 2), one gets a quadratic equation for $v_u$. Solving it gives:
 \beq  \label{6}
  \frac{v_u}{\sqrt{f_c}} = \frac{- 1 + \sqrt{1 + 4 \al_1 C_1 d}}{ 2 C_1 d},~~~~
  C_1=\frac{\al_2 \al_3 \ga \sqrt{f_c}}{d_{dg}}
 \eeq
which coincides with Eq. 5 in \cite{RuiMut18}. Multiplying both the numerator and denominator of the right-hand side of this equation with $1 + \sqrt{1 + 4 \al_1 C_1 d}$ and equating it to 1, a more instructive form of this equation follows:
 \beq \label{7}
  \frac{v_u}{\sqrt{f_c}} = \frac {2 \al_1 } {1 + \sqrt{1 + d/d_0^M}},~~~~
  d_0^M = \frac{1}{4 \al_1 C_1 }
 \eeq

This equation is in a form readily comparable to the size effect factor $\la$ proposed by ACI Committee 445, Fracture Mechanics. This factor will appear in the 2019 version of ACI code (ACI Standard 318), and reads:
 \beq \label{8}
  v_u = v_0 \la,~~~~\la = \frac 1 {\sqrt{1 + d/d_0}}
 \eeq
where $d_0$ is a constant, called the transitional size (in ACI taken as 10 in.). It needs to be pointed out that ACI also includes a horizontal cutoff at $v_u = 2 \sqrt{f'_c}$ (introduced by ACI-318E), which is not considered here because it is justified by statistics rather than mechanics, i.e., by an increasing width of the database scatter band on approach to smaller $d$, which causes the lower bound of the scatter band at small sizes to be almost independent of $d$.

An important feature is the asymptotics, which has been solidly established by quasibrittle fracture mechanics and is as follows:
 \beq \label{9}
  \mbox{For $d/d_0 \to 0:~v_u$ = constant;~~~
  For $d/d_0 \to \infty:~v_u \to (d/d_0)^{-1/2}$ }
 \eeq
These asymptotic properties are satisfied by both formulations. However, for the same transitional size, i.e., for $d_0^M = d_0$, the size effect curves of $\log(v_u/v_0)$ versus $\log(d/d_0)$ differ significantly and the large-size asymptote do not coincide; see Fig. \ref{f1}.

Varying the ratio $k = d_0^M / d_0$ stretches the size effect curve horizontally but does not change slope $-1/2$ in the log-log plot. Can the large size asymptotes be made to coincide by varying the $k$? To answer it, denote $q = d/d_0$, consider that the small-size asymptotes are matched, which occurs for $\al_1 \sqrt{f_c}= v_0$,
                    %8/6 I added /sqrt{f_c} to this eq.
and seek the value of $k$ for which the large size limit of the ratio of the size effect expressions in Eqs. (\ref{7}) and (\ref{8}) equals 1, i.e.,
 \beq \label{10}
  \lim_{d \to \infty} \frac{v_u^M}{v_u} =
  \lim_{d \to \infty} \frac{2\, \sqrt{1 + q}}{1 + \sqrt{1 + q/k}} =  1
 \eeq
The limit is $2 \sqrt{k}$ and, equating it to 1, one finds that both the small-size and large-size asymptotes get matched if
 \beq \label{11}
  k = 1/4~~~~\mbox{or}~~~~d_0^M = d_0 / 4
 \eeq

        %AD -Prof. Bazant The below paragraph seems unclear to me. In Modified Compr Field Theory paper in the second paragh of page 617 explains this situation.
It may also be noted that Eq. (\ref{1}) with (\ref{2}) was previously used in MCFT  \cite{CollinsVechio, Toronto} without the dependence of $w$ on $\eps$ and was in this form adopted for the Level I Approximation in Model Code 2010. In that case, the size effect curve ended with asymptotic slope $-1$ instead of $-1/2$, which is thermodynamically impossible. Muttoni and Ruiz's \cite{MutRui08} artificial modification that added the assumed dependence on $\eps$ made $w$ proportional to $v_u$. This then led to the quadratic equation for $v_u$, Eq. (\ref{5}), and thus changed the asymptotic slope from the (thermodynamically impossible) value $-1$ to the value $-1/2$ dictated by fracture mechanics.

%AD starts
\subsection*{\large Deficiencies of CSCT Revealed by FE Simulations of Beam Shear Failure}

Certain key aspects of failure, such as finding where the energy needed for fracture is coming from and where it is dissipated, are virtually impossible to observe in experiments directly. However, they can be revealed by extending experimental results with a realistic computer model. Microplane constitutive model M7 \cite{CanBaz13a, CanBaz13b} for softening damage in concrete, combined in FE element analysis with the crack band model \cite{BazOh83, BazCerWie}, is such a model, as proven in many previous studies \cite[e.g.]{DonBaz17, VorelBaz14, RasBaz18} and also verified in Appendix 2 (which gives more information on the FE analysis). We will pursue the FE approach now, examining not only the energy flow but also other features important to understand the shear strength, such as the stress distributions across damage zones and along cohesive cracks, stress redistributions due to fracture and the overall response of structures of different sizes and shapes. For the size effect analysis to make sense, the modes of failure, and particularly the shapes of dominant cracks in geometrically similar beams of different sizes must also be geometrically similar. This fact has been experimentally best documented by the tests of Syroka-Korol and Tejchman \cite{Korol14}, as shown in Fig. 7 of \cite{YuLe--16}.

%If a highly realistic damage constitutive law is available, then the finite element (FE) analysis with the crack band model, calibrated by limited test results, has been shown to yield realistic extensions of the experimental information to detect various hard-to-measure features of fracturing structures, such as the stress distributions in damage zones and along cohesive cracks, and the overall response of structures of similar but different shapes and sizes. It has been shown  that the microplane model, and particularly its latest version M7 \cite{CanBaz13a, CanBaz13b}, is such a constitutive law. Indeed, comparisons with diverse experiments have shown that, when the free parameters of M7 are calibrated by one experiments, one gets good predictions of other experiments with the same concrete; %(Fig. \ref{f3});      %9/11 moved to Appendix
%see the verification examples in Appendix 2, which also give further information on the FE analysis.

Shear tests of geometrically similar RC beams of different sizes, without stirrups, have been simulated with FE program ABAQUS. The microplane damage constitutive model M7 has been implemented in user's subroutine VUMAT.
   %M7 is the latest and most realistic in a series of microplane models for concrete developed at Northwestern University.
       %9/11 Dr. D:  I will now reference figures by numbers in the last pdf file. Please rearrange and renumber the figures in the sequence I will quote them.
The carefully conducted experiments of beam shear failure performed in Collins' lab at the University of Toronto \cite{Toronto} are chosen to calibrate the FE element code with constitutive model M7, as detailed in Appendix 2.

Fig. \ref{f2}a shows, in relative coordinates scaled with the beam size, the FE meshes for nearly similar beams of two sizes, which are $d=110$ mm and 924 mm (this gives the size ratio 8.41\,:\,1). The FE size is the same for all beam sizes, so as to avoid dealing with spurious mesh size sensitivity. The blackened elements in Fig. \ref{f2} show, for maximum load $P_{max}$, the locations of the highest longitudinal strains in the element. Note that the band of blackened elements runs faithfully along the upper side of main crack (the width of the blackened band is the finite element width and has nothing to do with the crack width).

\subsubsection*{\normalsize a) Localizing stress distributions}

Fig. \ref{f2}a shows, to the right of the beams, the distributions of longitudinal normal stress $\sig_x$ across the height of the ligament (understood as the zone between the tip of the dominant diagonal shear crack at $P_{max}$ and the top face). These distributions (which are similar to those in \cite{BazYu05b} and in Fig. 11 of \cite{YuLe--16}) demonstrate how the stresses localize near the top face as the size $d$ is increased. It is evident that, in the smallest beam, the material strength across the ligament is utilized almost in full (for plastic behavior it would be mobilized fully). In the largest beam, by contrast, the material strength is, at $P_{max}$, localized within only a small portion of the ligament. This localization provides an intuitive explanation of the size effect (see also \cite{BazYu05b,BazYu05a,aci446}). What do these distributions have in common with the crack opening at depth $0.6d$, and generally with Hypotheses 3 and 5? Nothing.

\subsubsection*{\normalsize b) Cross-crack stress transmission}

Fig. \ref{f2}a reveals, for beams of two sizes, another relevant feature impossible to measure directly---the distribution of the vertical stress component transmitted, at $P_{max}$, across the crack. The numbers at the vertical arrows represent, as a percentage of tensile strength of concrete, the vertical stress transmitted at $P_{max}$ across the diagonal shear crack due to the aggregate interlock or cohesive softening. Here it is important to notice that, except near the crack tip, the percentages are quite low, and that they decrease markedly with the beam size.

Especially note in Fig. \ref{f2}a the vertical forces $V_{crack}$ indicated as percentages of the total shear force, $V_c = P_{max}$, obtained by integrating the vertical cross-crack stress components along the whole crack length. They represent for the beams of small and large sizes, respectively, only 18\% and 2.9\% of the total shear force $V_c$. If the vertical force transmitted at $P_{max}$ across the crack is so small, how could the crack width, $w$, play any significant role? It could not. Therefore, Hypotheses 2--5 are unjustified, unrealistic.

   %9/11 Dr. D: Fig. 5b is too cluttered and cannot be presented as is. I suggest showing only a few of the principal stress vectors, of length proportional to magnitude, in dark color, to be visible. Isn't there a program to trace the principal stress trajectories, whose spacing inverse indicates the magnitude? ZPB
Fig. \ref{f2}b shows the vectors of minimum principal compressive stresses (maximum in magnitude). They confirm that the load just before the failure (i.e., at $P_{max}$) is transmitted mainly by a strip of concrete along the top side of the crack. In the sense of the strut-and-tie model, this strip represents what is called the "compression strut". The fact that these vectors generally do not cross the crack means that the force transmitted across the crack at $P_{max}$ is negligible. This again contradicts Hypotheses 2--5.

%AD-I edited the below sec
\subsubsection*{\normalsize c) Scenario of energy release and dissipation}

Most relevant for fracture is the energy picture, shown in Fig. \ref{f3}. Fracture dissipates energy, and that energy must come from somewhere. At controlled displacement, it must come solely from the release of potential energy (i.e., strain energy) from the structure.
  %, which is a sum of the strain energy and the potential energy of load.
As proposed by Griffith in 1921 \cite{BazPla98}, this release is a central tenet of fracture mechanics of all types, including quasibrittle fracture. So we calculate, for all the integration points of all the elements, the density of strain energy released at unloading:
 \beq \label{U}
  \bar \Pi = \mbox{$\frac 1 2$}\ \ssig^T \CC \ssig
 \eeq
which is an important quantity that cannot be directly measured; $\CC$ is the $6\times 6$ matrix of elastic compliances for unloading (i.e., inverse of the elastic moduli matrix), and $\ssig$ is the $6 \times 1$ column matrix of stress components, as affected by distributed fracturing (for simplicity, the unloading stiffness is considered the same as the initial elastic stiffness). Then, summing over all the integration points, we calculate the energy change between two states, in this case
 \beq  \label{Del U}
  \Del \Pi = \Pi_{99} - \Pi_{00}
 \eeq
which is from the value $\Pi_{99}$ for the prepeak state at load $P = 0.99 P_{max}$ to the value $\Pi_{00}$ for the postpeak state at which the load has been reduced to 0 (other states could be chosen, too; but for states close to each other the changes would have high numerical scatter).

The energy difference $\Del \Pi$ represents the values of energy density release at each integration point of the finite elements. From all these values, one can compute the contour plots of the zones of energy release density. These zones are shown in relative coordinates in Fig. \ref{f3}, where the zone of maximum density is in dark. Note that, in actual coordinates, the dark band for the largest beam would be much wider than for the smallest beam.

The dark band of the energy release, which may be imagined to represent the `compression strut' of strut-and-tie model. The strut is located wholy above the main crack. On top of the strut, the energy release comes from damaging concrete, and in the rest of strut from the unloading of undamaged concrete. The main diagonal shear crack does not contribute to the dominant energy changes, especially not for large beams. So how can it play a significant role in controlling the failure load? Again this contradicts Hypotheses 3 and 5.

The essential idea of the SEL is that the energy release from the structure is a sum of two parts, $\Pi = \Pi_1 + \Pi_2$, where $\Pi_2$ is the strain energy released by unloading from the undamaged part of the structures whose volume increases (for geometric scaling) {\em quadratically} with the structures size ($d$ in our case), and $\Pi_1$ is the strain energy released by unloading from the damaged part (traveled by the fracture process zone) whose volume increases {\em linearly} with the structure size, whereas $W$, the energy dissipated (which must be equal to $\Pi$) increases always {\em linearly} with structure size. The ratio of the quadratic to linear increases immediately indicates that, for small enough sizes, the quadratic part, $\Pi_2$, must be {\em negligible} compared to the linear part, $\Pi_1$, while for large enough sizes it must be {\em dominant}.

In our problem, further evaluation of the FE results could show that the total energy dissipation by fracturing damage increases roughly {\em linearly} with the shear crack length, which is proportional to beam size $d$, while the energy released from the undamaged part of the dark band (or compression strut) on the side of the main crack increases roughly {\em quadratically} with $d$ because not only the length but also the width of the dark band (or the strut) in Fig. \ref{f3} increases roughly in proportion to $d$.
Thus the mismatch of linear and quadratic increase is the ultimate source of transition in the size effect curve for beam shear.

%9/11 Dr. D: It might help to plot the red bands for the smallest and largest beam from Fig. 4b in actual rather then relative coordinates. I expect the red band to be much fatter for the large sizes. ZPB

The aforementioned energy derivation of the size effect law, SEL, is, in fact, much simpler than that of CSCT (cf. Appendix 3, and also the 1984 study \cite{Baz84} in which the SEL was formulated). Energy conservation and dimensional analysis is the essence of a fundamental but simple derivation of the SEL as given in Eqs. 5--7 of \cite{YuLe--16} and summarized in Appendix 3.

\subsubsection*{\normalsize d) Compatibility with modernized strut-and-tie model}

%The red (or dark) strips of high energy release on the side shear crack, seen in Fig. 4b, actually represent the ``compression struts" of the popular strut-and-tie model (originally called M\" orsch's truss analogy \cite{Baz07}). This model

The strut-and-tie model (originally called M\" orsch's truss analogy) is generally considered to provide good estimates of limit loads of concrete structures. In the classical form, however, the strut-and-tie model misses the size effect. In \cite{Baz97} it was shown how the strut-and-tie model could be modernized by calculating the balance of energy dissipated by compression-shear crushing on top of the compression strut, with the energy release from the intact and damaging parts of the compression strut.
                           %Such modernized strut-and-tie mode is perfectly compatible with the SEL. but incompatible with the CSCT.

Recently it has become widely accepted that the strut-and-tie model must be modernized by incorporating the size effect into the compression struts. In view of the foregoing observations about the energy release zone and the energy dissipation zone on top of the strut, such a modernization is simple, obvious and logical---introduce the size effect into the compression strut (some researchers work on it alerady).

The concept of a modernized ``compression strut" exhibiting a size effect is in concert with the SEL, agrees with the present analysis (Fig. \ref{f4}) as well as with the conclusions in \cite{BazYu05b, Baz97}. Doesn't this agreement invalidate the hypotheses of CSCT? It certainly does. It must be concluded that the CSCT is incompatible with the modernized strut-and-tie model, whereas the SEL is.

\subsection*{\large Unfounded and Scientifically Dubious Aspects of the Hypotheses}  % Underpinning CSCT}

\hskip 5 mm {\bf Re Hypothesis 1.} According to the FE simulations,
%AD shown in Fig. \ref{f4}a and b, I deleted this words
the crack width $w$ is highly variable along the crack length. Which opening, $w$, at which location and beam size, produces the cohesive stress that matters for the ultimate shear force $v_u$? There is no answer.

{\bf Re Hypothesis 2.} A basic concept of the cohesive crack model, which includes cohesion due to the aggregate interlock, is the decrease of crack-bridging cohesive stress $\sig_c$ with increasing crack opening width $w$. This is a property studied for concrete in great detail since 1980 \cite{Walraven81}. Already by 1990 it became clear that Eq. 1, with $w$ appearing linearly in the denominator, agrees neither with the experimental evidence on postpeak softening of cracks in concrete, nor with FE simulations. It is now generally accepted that the cohesive softening curve is approximately bilinear, with a steep initial drop followed by a very long tail \cite[e.g.]{HooBaz14, HooBaz13}.

If the denominator of Eq. 1 were anything but linear in $w$, substitution of Eq. (2) into Eq. (1) would not yield for CSCT a quadratic equation for $w$ (Eq. 5 in \cite{MutRui08}), and then the large size asymptote of size effect would not have slope $-1/2$ in log-scale.

So the only justification of Eq. 1 is convenience---to manipulate the derivation so as to obtain an asymptote of slope $-1/2$, which is by now a widely accepted fact.

{\bf Re Hypothesis 3.} Why should the crack width, and thus the ultimate load and the size effect, be determined by some strain, $\eps$, at some specific location? That is pure fiction, and is impossible in fracture mechanics.

Can one identify in the present FE results any characteristic strain controlling the ultimate load? Certainly not. What matters for fracture is the release of stored strain energy from the structure and, in the case of cohesive (or quasibrittle) fracture, also the tensile strength of material. Certainly not any strain.

{\bf Re Hypothesis 4.} Why should the reference strain, $\eps$, be taken at distance $d/2$ from the concentrated load, and why at depth $0.6d$ from the top face? This is a mystery justified only be convenience, to get results not in gross violation of the experimental evidence.

{\bf Re Hypothesis 5.} Why should the reference strain, $\eps$, be calculated according to linear elasticity if, at ultimate load, the concrete is behaves highly nonlinearly, due to cracking damage? This is fiction.

And why should the strain analysis use the classical beam bending theory based on the Bernoulli-Navier hypothesis, which applies only in the flexure of sufficiently slender beams? Fiction again.

The FE analysis shows the initially plane cross sections at ultimate load to be highly warped. Formulas to calculate $\eps$ based on inelastic behavior would, of course, be awfully complicated. But why to bother with calculating this strain at all? It is not the cause of failure. The real cause of quasibrittle failures and size effect is the overall energy release from the structure.

{\bf Re Hypothesis 6.} The overall spirit of the size effect calculation in CSCT (as well as MCFT) is to avoid fracture mechanics and replace it by some sort of simple linear elastic beam analysis. This analysis is just an artifice, aimed to provide a semblance of some logic.

\subsection*{\large Can CSCT Be Extended to Designs Outside the Range of Existing Tests?}

The bulk of the existing tests of size effect in beam shear strength involved a relatively limited range of geometrical situations, in terms of shear span ratio and reinforcement types, and did not include continuous beams. It is questionable whether the CSCT, beginning with the elastic strain at $0.6d$ depth, could be applied to such situations.

On the other hand, the size effect factor $\lambda$, Eq. (\ref{8}) based on SEL, is, in principle, applicable to all quasibrittle failures (geometrically similar for different sizes), in which a long stable crack develops prior to reaching the maximum load and an unlimited postpeak plastic plateau is lacking. Generally, it suffices to multiply with $\lambda$ the limit analysis formula for the strength contribution of concrete that works for small beam sizes. The only parameter that needs to be estimated is the transitional size $d_0$, although one can assume that it varies negligibly within the normal range of geometries.

\subsection*{\large Shouldn't the Design Code Heed the Ideals of Simplicity and Generality?}

The main problem with Eq. (\ref{6}) or (\ref{7}) of CSCT is not that it would be unsafe to a significant degree. It is not.

The problem is that the fictitious derivation obfuscates the mechanics of failure and is much more complicated than necessary to obtain a realistic size effect prediction. Just compare the derivation discussed above with the general derivation of size effect law in Eqs. 4 - 7 in \cite{YuLe--16}, based solely on the energy conservation and dimensional analysis (as sumarized in Appendix 3), or to the original 1984 derivation in \cite{Baz84}, based on approximation of energy release in presence of a localization limiter, the characteristic size of the fracture process zone.

These, as well as several other, derivations of the size effect law are much simpler, and are based on only three hypotheses---the relevance of energy release, the geometrical similarity of dominant cracks for different sizes, and the approximate size independence of the fracture process zone (which represents a characteristic length as a material property). These hypotheses are both obvious, generally accepted and generally applicable to many types of structures and materials. These are all the quasibrittle materials which, aside from concrete and mortar, also include the fiber-polymer composites, tough ceramics, sea ice, many rocks, stiff soils, masonry, wood, etc. They all exhibit the same kind of size effect on nominal structural strength.

So why should the size effect in concrete be different? Concrete shear failure is not as exceptional as the derivation of CSCT suggests. Rather it is just one manifestations of a typical size effect exhibited by many materials and structures. So why should the beam shear need a special derivation, not applicable to all these similar situations?
     %is one application of the size effect  If they fail after large stable crack growth, as %in beam shear, they all follow the same energetic size effect law, as amply demonstrated %over the last three decades in virtually hundreds of studies.
Is it not strange that the purported derivation underlying size effect of the CSCT (or the MCFT and Model Code 2010) cannot be transplanted to other quasibrittle materials? Why should concrete, and the shear of beams, be so {\em unique}?

Besides, a formulation based on the energy release concept of fracture mechanics (based on the work of Ballarini at Northwestern in the 1980s) has already been used for a long time in most design codes, to predict the shear failure in the pullout of anchors from concrete walls, including the size effect. How come that, for one type of shear failure of concrete, the fracture mechanics basis of size effect is accepted in the Model Code, while for another type of shear failure it is not?

\subsection*{\large Comments on the Analogous Problem of Size Effect on Punching Shear Strength of Slabs}

As demonstrated in \cite{BazCao, DonBaz17}, the punching shear strength of slabs also follows the size effect law derived from energy release (and the SEL size effect factor will appear in the 2019 version of the ACI code). Nevertheless, an alternative size effect calculation based on CSCT, resting on elementary mechanics of bending, was incorporated into Model Code 2010. Muttoni et al. \cite{MutRui13} adapted their CSCT to punching with the modification that a certain reference slab rotation is used instead of the reference strain. For punching they thus obtain a size effect that ends with the asymptotic slope of $-0.4$ (instead of $-0.5$), which is not correct but at least does not violate the second law of thermodynamics. The hypotheses in the derivation are again unjustified and fictitious.

\subsection*{\large Conclusions}

  \be \setlength{\itemsep}{-1.3mm}
  \ii
The shear size effect of the Muttoni et al.s CSCT exhibits the correct small-size and large-size asymptotic behaviors and can fit the size effect test data almost as well as the energetic size effect law (SEL).
  \ii
The size effect of Muttoni et al.'s CSCT is based on a simplistic derivation contrived to give a semblance of logical support in mechanics. The calculation of CSCT (as well as MCFT) is misleading. It is a `cookbook' procedure that obfuscates the mechanics, while none exists of shear failure. It would have to be taught to students as a dogma.
  %  \ii
  %From the practical viewpoint, the error of the CSCT is not a big problem. k,mn T\  he  obfuscation is.
           %  \ii
           %The CSCT involves a cannily manipulated calculation procedure that delivers, despite its %erroneous hypotheses, a size effect curve with the correct asymptotes, reasonably close %to the widely accepted SEL.
  \ii
The CSCT is shown to rest on six implied hypotheses. They are all physically unjustifiable. They involve application of the classical one-dimensional elastic beam bending theory to what is a multidimensional nonlinear problem of fracture mechanics. The same can be said about the hypotheses implied by the MCFT and Model Code 2010.
  \ii
Finite element simulations with the M7 constitutive model, calibrated and verified here by the classical Toronto tests, extend the measured data by showing that, within the ligament between the tip of the main crack and the beam top, the stress profile is nearly uniform for small beams and rather localized for large beams. This means that the concrete strength is mobilized almost fully for small beams and only partly for large beams.
  \ii
According to finite element results, the energy dissipation during fracture comes mainly from a highly stressed band on the side of the main shear crack and from a small damage zone above the tip of that crack. The main shear crack dissipates during failure almost no energy and thus, contrary to CSCT, its opening width is not what controls failure. This observation suffices to invalidate the CSCT (as well as MCFT).
  \ii
Because of the lack of support in mechanics, the CSCT cannot be trusted for extensions to designs outside the range of bulk of the existing size effect test data, which include different reinforcement types and shear spans, or continuous beams.
  \ii
The CSCT, as well as MCFT, is incompatible with the strut-and-tie model, while the SEL is, provided that the size effect is incorporated into the compression strut (which is a currently pursued goal).
 \ii
The size effect of MCFT, incorporated into Model Code 2010, shows major deviations from the SEL. It has an incorrect large-size asymptote that is thermodynamically inadmissible and, consequently, it mispredicts the size effect in large beams.
  \ee
               %It would be much simpler to apply the SEL factor to the classical calculation of shear strength based on limit analysis.
%  \ii
%The MCFT is as simple as applying the SEL factor but grossly overestimates the %size effect for very large beams, while mildly underestimating the size effect %in the mid-size range.

\vv \no {\bf Acknowledgment:}\ Partial funding under NSF Grant No. CMMI-
1439960 to Northwestern University is gratefully acknowledged.

\subsection*{\large Appendix 1: Deficiencies and Problems of MCFT and of Model Code 2010}

Although the MCFT, featured in the existing Model Code 2010, is not the focus of this study, some points are worth noting, for comparative purposes. The  MCFT has more serious deficiencies than the CSCT. The Level I Approximation of Model Code 2010 consists of Eq. (1) in which the value of $w d_{dg}$ (and thus also $\eps$) is not variable but is fixed.
                      %(as a minor difference, $z$ is used instead of $d$).
This gives $\al_2 w / d_{dg} d = 1.25 z$. Consequently, the large-size asymptotic behavior is
 \beq \label{-1}
  \mbox{for}~ d \to \infty:~~~v_u \to \frac{\mbox{constant}} d
 \eeq
Such asymptotic behavior is not supported experimentally. It is, in fact,  thermodynamically impossible. Extrapolation to large sizes would severely exaggerate the size effect. At the same time, since the transition from the small-size (horizontal) to the large-size asymptote is sharper and narrower than it is for the energetic size effect law, Eq. (\ref{-1}) underestimates the size effect in the mid-size range if the size effect is fitted to the same small-size data.

It may be noted that an equation of the same form as (\ref{-1}) was proposed in \cite{BenCol06}. It found its way into the 2004 Canadian CSA A23.3 shear design provisions. A similar critique applies.
   %It was claimed that the size effect was explained by: 1) increase of crack width $w$ with structure size (or $d$), and 2) proportionality of $w$ to longitudinal strain $\eps_x$ calculated by elastic bending theory.

The Level II approximation of Model Code 2010 (or MCFT, Eq. 4a in \cite{MC-Background}) is written as
 \beq  \label{m1}
  \frac{v_u}{\sqrt{f_c}} = k_v,~~~~
  k_v = \frac{0.4}{1 + 1500 \eps}\ \frac{1300}{1000 + d_{dg} z}
 \eeq
which may be rewritten as
 \beq  \label{m2}
  k_v = \frac{c_1}{ (1 + c_2 k_v )( c_e + c_4 z) }
 \eeq
in which $c_1 = 520,~c_2 = 1500 c \sqrt{f_c},~c_3 = 1000,~c_4 = k_{dg}$ are constants.
  %This can be algebraically rearranged into a quadratic equation for the size effect factor $k_v$:
  %\beq \label{m3}
  %(c_5 + c_6 z) k_v^2 + (c_3 + c_4 z) k_v - c_1 = 0
  %\eeq
  %in which $c_5 = c_2 c_3, c_6 = c_2 c_4$ are constants. Note that depth $z$ %appears in this equation in a different way than $d$ does in Eq. (\ref{5'}), and this changes the asymptotic slope.
Although this equation leads to a quadratic equation for $k_v$ (different from Eq. (\ref{1}) ), the asymptotic slope $k_v$ for $z \to \infty$ may be more directly determined by replacing $k_v$ with a new variable $F$ such that
 \beq \label{m3}
  k_v = F/z
 \eeq
Eq. (\ref{m2}) may then be rearranged as
 \beq \label{m4}
  F \left( 1 + c_2 \frac F z \right)\left( \frac{c_4} z + c_4 \right) = c_1
 \eeq
Now, assuming that $F$ is a constant, the limit of this equation for $z \to \infty$ is $F(1 + 0)(0 + c_4) = c_1$, i.e. $F c_4 = c_1$ or $F = c_1 / c_4$. This confirms our assumed constancy of $F$ to have been correct and that $k_v = (c_1/c_4) / z$, or
 \beq
  \mbox{for~~ $z \to \infty$:}~~~~k_v\ = \ \frac{\mbox{constant}} z
 \eeq
Such asymptotic behavior of Level II approximation of MCFT and of Model Code 2010 is, of course, also thermodynamically impossible, and thus untenable (same as for Level I). It also reveals a lack of scientific basis.
           %AD A comparison of the MC 2010 LoA I and II with SEL is demonstrated in Fig \ref{f5} with using and extended database \cite{database}-I deleted this part

Most of the hypotheses of CSCT also apply to MCFT, and the previous criticisms need not be repeated.

\subsection*{\large Appendix 2: Examples Verifying Realistic Performance of Microplane Model M7}

The credibility of the foregoing FE analysis with model M7 depends on comparisons with experiments. Model M7 (whose coding can be freely downloaded from www.civil.northwestern.edu/people/bazant/ ), appears to be the only one that can match all types of material tests of concrete, as shown in \cite{CanBaz13b}. M7 calibrated by a part of the data set on various structural tests was shown to predict correctly the rest of the data set \cite[e.g.]{CanBaz13b, DonBaz17, YuLe--16, BazYu05b}. Here we show how well the M7 fits the strength and size effect of the three trusted experiments of beam shear failure and size effect \cite{Toronto, Korol14, Walraven} (similar demonstrations were also made in \cite{BazYu05b}).

Fig. \ref{f4} shows the fitting of these tests, used for calibration of M7. In Fig. \ref{f4}a, four-point-bend specimens of 4 different sizes (with only approximate geometric similarity) are simulated by finite elements (FE) using M7. The smallest and the largest effective depths are $d$ = 110 mm and 925 mm. The flexural reinforcement ratio slightly varies from 0.76 \% to 0.91 \%. The shear span ratio, $a/d$, is 3. The mesh for concrete uses 3D continuum hexahedral elements of size 12.5mm, which are kept the same for all sizes in order to avoid dealing with spurious mesh sensitivity due to localization of softening damage. The reinforcement is modeled with 2-node linear beam elements attached at nodes to the elements of concrete. The smallest FE system has 1457 nodes and 990 elements, while the largest one has 78,895 nodes and 58,347 elements.

Fig. \ref{f4}b shows the verification and calibration results for tests conducted in \cite{Korol14}. Beams of three sizes are tested with a 4-point bending load configuration. The effective depths of beams are 160, 360 and 750mm. The reinforcement ratio is 1.0\% and the aspect ratio, $a/d$, is 3. The maximum aggregate size is 10mm. The element size is 20mm, for all the sizes. 3D continuum elements with reduced integration are used in explicit (dynamic) analysis.

Fig. \ref{f4}c demonstrates the results for Walraven tests \cite{Walraven} for normal weight concrete. Three different sizes with effective depths of  125, 420 and 720 mm with four point bending configuration are modeled and calculated with 3D hex elements. The concrete strength is 34.2 MPa and the reinforcement ratio slightly varies from 0.75\% to 0.83\%. Fig. \ref{f4} also shows the the differences in the ultimate shear strength predictions of CSCT for the used tests. The variation of the secondary parameters other than the size could result in very high discrepancies for ultimate strengths between the size effect law and CSCT. For example in Fig. \ref{f4}c, the difference reaches to 29 \% for small sizes and doubles the difference for $d \to \infty$.

\subsection*{\large Appendix 3.\ For Comparison: General Derivation of SEL from Energy Conservation and Dimensional Analysis}

The total release of (complementary) strain energy $\Pi$ caused by fracture is a function of both 1) the length $a$ of the fracture (or crack band) at maximum load, and 2) the area of the zone damaged by fracturing, which is $w_c a$, where $w_c = n d_a$ = material constant = width of crack band swept by fracture process zone width during propagation of the main crack, $d_a$ = maximum aggregate size, and $n$ = 2 to 3. Parameters $a$ and $w_c a$ are not dimensionless, but can appear only as dimensionless parameters, which may be taken as $\al_1 = a/D$ and $\al_2 = w_c a/D^2$, where $D$ =total beam depth $h$ or depth $d$. According to the Buckingham theorem of dimensional analysis, the total strain energy release must have the form:
 \beq \label{a4}
  \Pi = \frac 1 {2E} \left( \frac P {bD} \right)^2 b D^2 f(\al_1, \al_2)
 \eeq
where $b_w$ = beam width. In the case of geometrically similar beams of different sizes, $f$ is a smooth function independent of $D$. The energy conservation during crack propagation requires that $\pa \Pi / \pa a = G_f b$, where $G_f$ = critical value of energy release rate. Now note that
 \beq \label{a5}
  \frac{\pa f}{\pa a} = \frac{\pa f}{\pa \al_1}\, \frac{\pa \al_1}{\pa a}
                     +  \frac{\pa f}{\pa \al_2}\, \frac{\pa \al_2}{\pa a}
 \eeq
where $\pa \al_1 /\pa a = 1/D$ and $\pa \al_2 /\pa a = w_c /D^2$, and consider the first two linear terms of a Taylor series expansion $f(\al_1, \al_2) \approx f(0, 0) + f_1 \al_1 + f_2 \al_2$, where $f_1 = \pa f /\pa \al_1$, $f_2 = \pa f/ \pa \al_2$. This leads to the equation
 \beq  \label{a6}
  \left( \frac{f_1} D + \frac{f_2 w_c}{D^2} \right) \frac{P^2}{2 b E}
  = G_f b
 \eeq
After rearrangement, and using the notation $v_u = P/b_w D$ = average (or nominal) shear strength due to concrete, Eq. (\ref{a6}) yields the deterministic (or energetic) size effect of ACI-446, with the size effect factor $\la$ given by Eq. (\ref{8}), in which $D_0 = w_c f_2/f_1$ = constant (independent of size $D$, characterizing structure geometry). Q.E.D.

The hypotheses underpinning this derivations are two: 1) The size, $w_c$ (width or length), of the fracture process zone at the front of dominant crack is constant (a material property), and 2) the failures are geometrically similar (this similarity is not listed here among the hypotheses of CSCT but is tacitly implied). Energy conservation is not a hypothesis but a physical fact. Neither is Eq. \ref{a4}, which is dictated by dimensional analysis. Note that the fracture mechanics had not to be specifically invoked in this derivation, although energy balance is the basis of fracture mechanics.

\listoffigures

\clearpage
%1
\bfi
\centering
\includegraphics[width = 65mm, height=50mm]{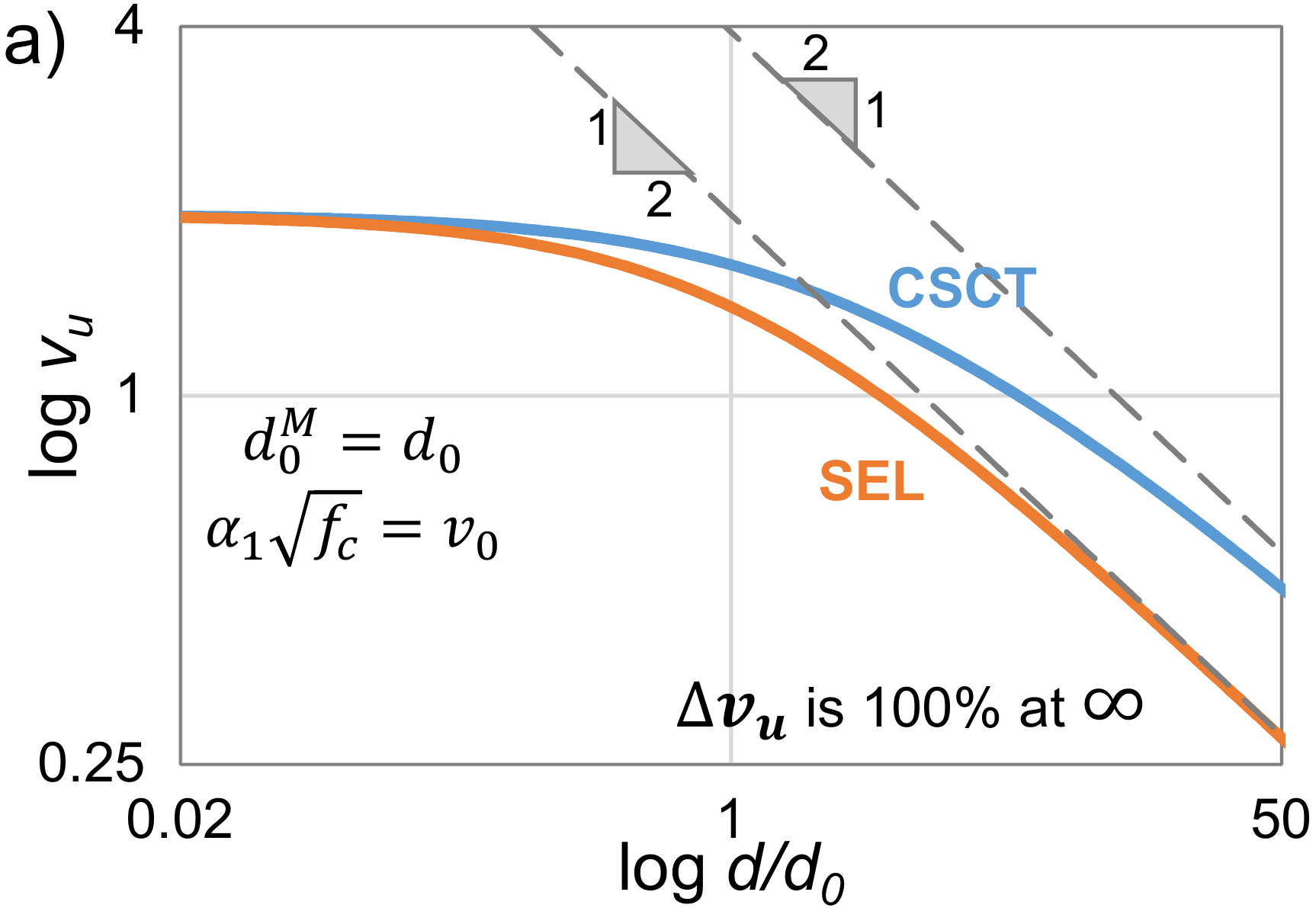}
\includegraphics[width = 65mm, height=50mm]{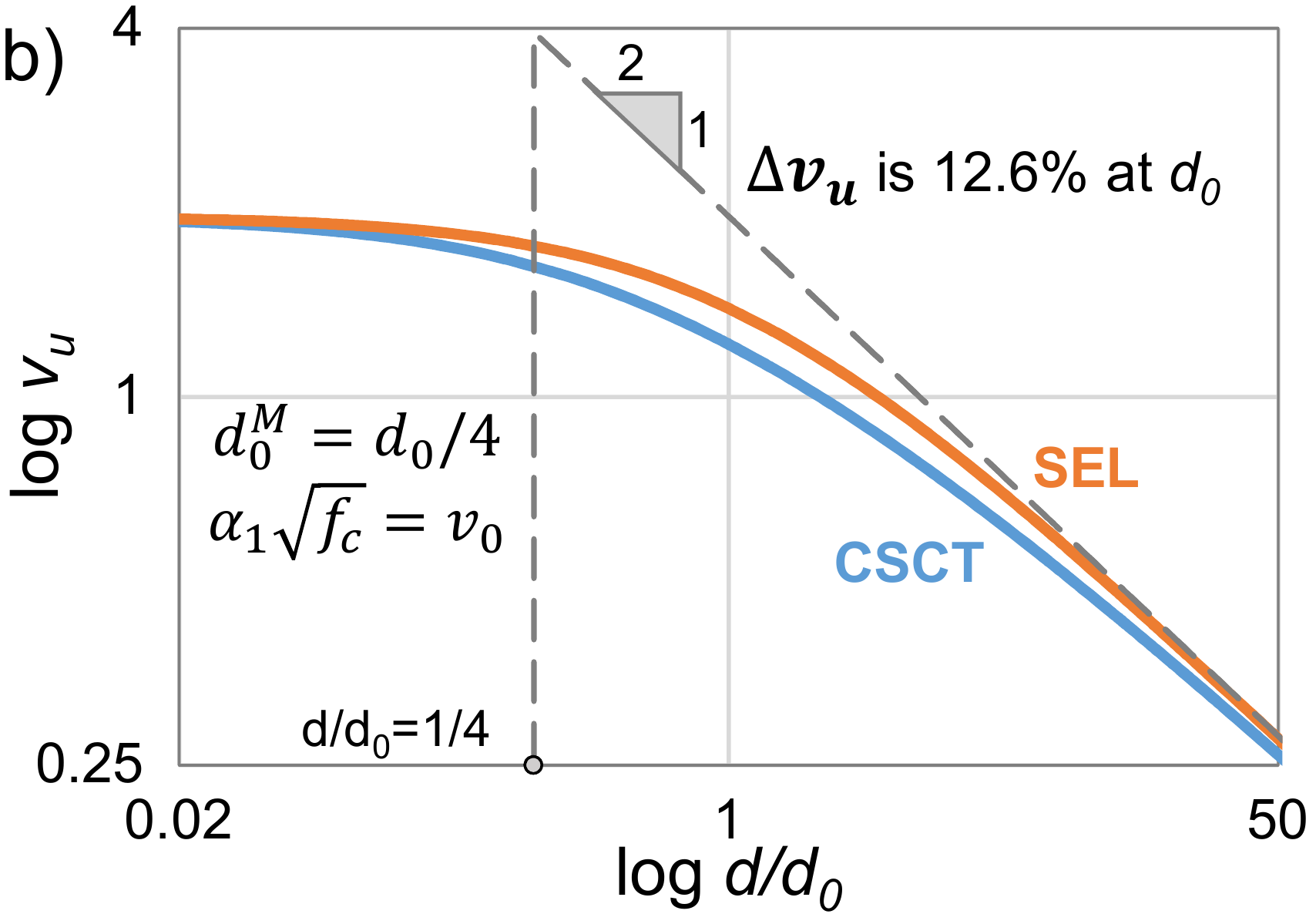}
\includegraphics[width = 65mm, height=50mm]{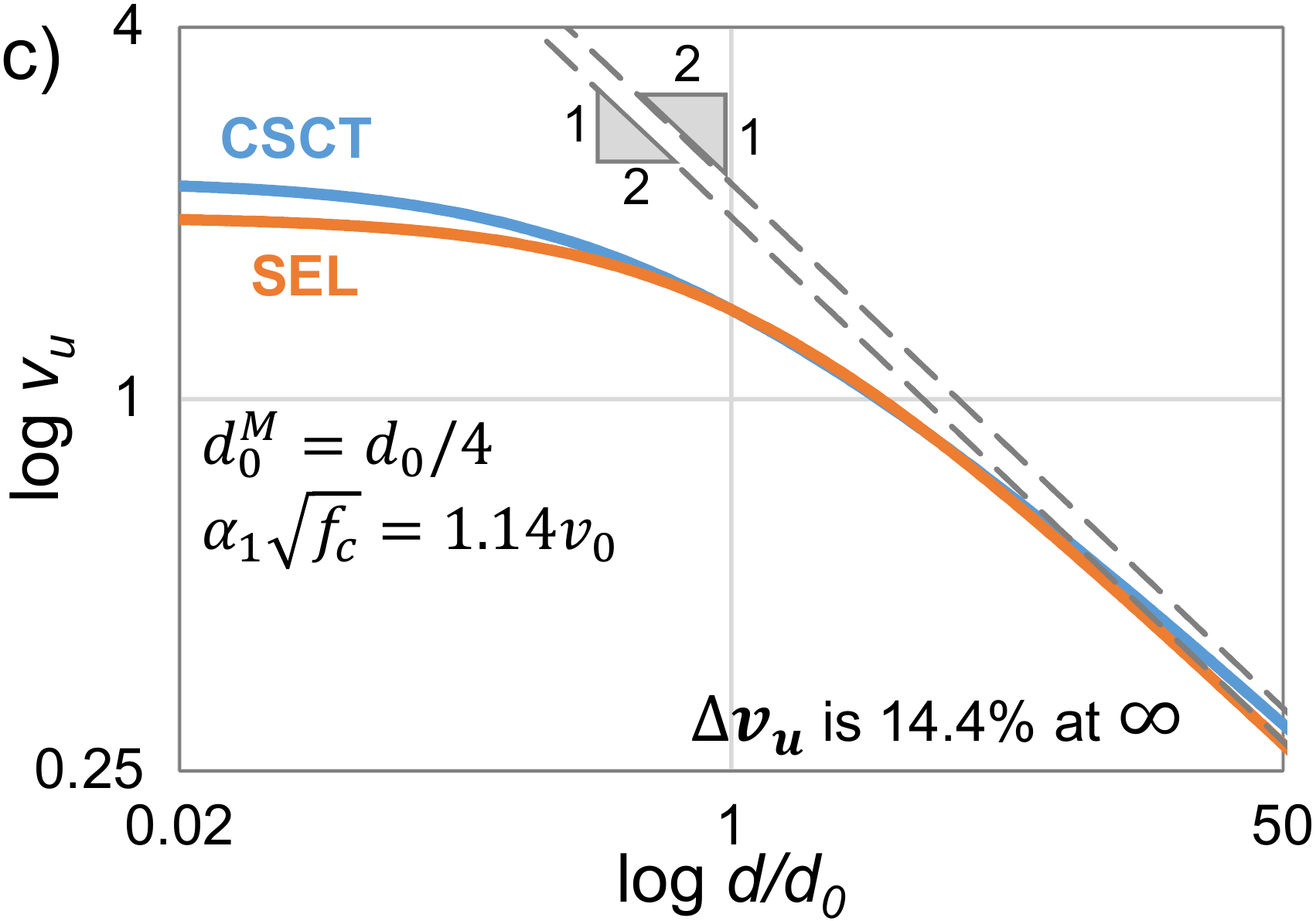}
\caption{\label{f1} \sf Size effect differences between ACI-446 \cite{aci446} \underline{S}ize  \underline{E}ffect \underline{L}aw and Model Code 2010 for beam shear.}
\efi

%2
\bfi
\centering
\includegraphics[width = 75mm, height=60mm]{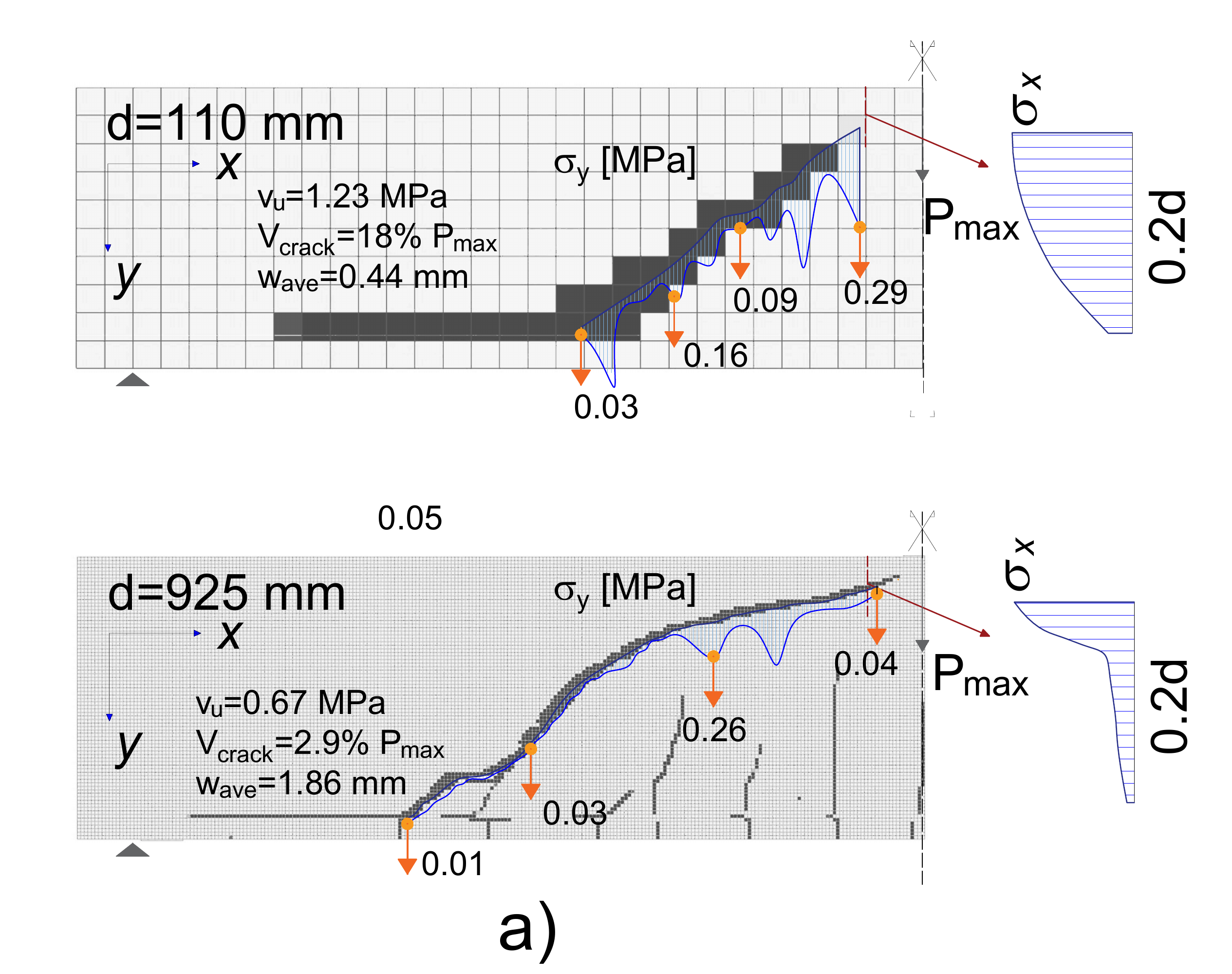}
\includegraphics[width = 62mm, height=60mm]{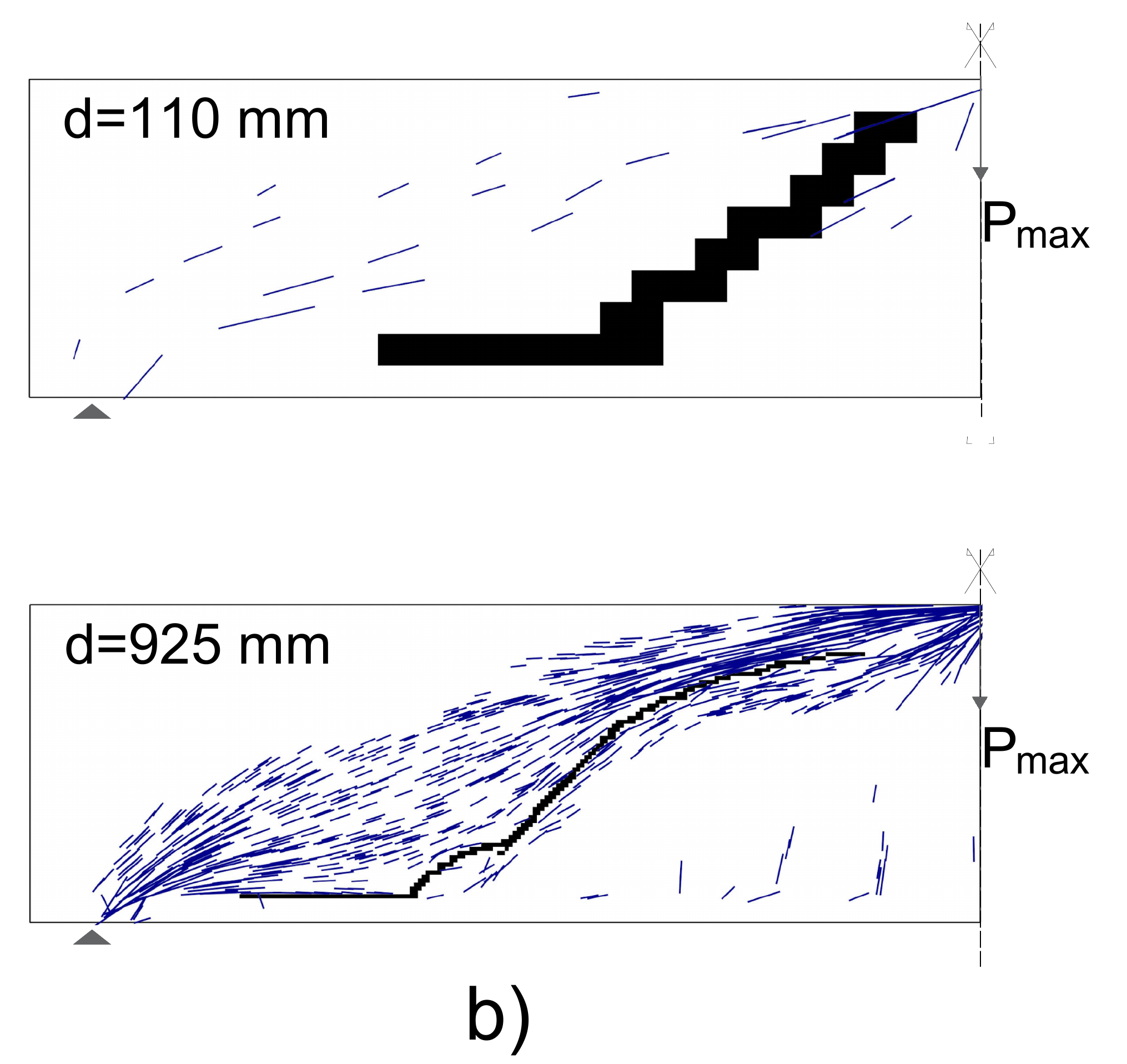}
\caption{\label{f2} \sf a) Longitudinal stress variation across ligament above the main crack tip (on the right of beam) and variation of vertical stress component along the diagonal crack, b) the vectors of minimum principal compressive stresses (maximum in magnitude) calculated for Toronto test beams \cite{Toronto}.}
\efi

%3
\bfi
\centering
\includegraphics[width = 85mm, height=52mm]{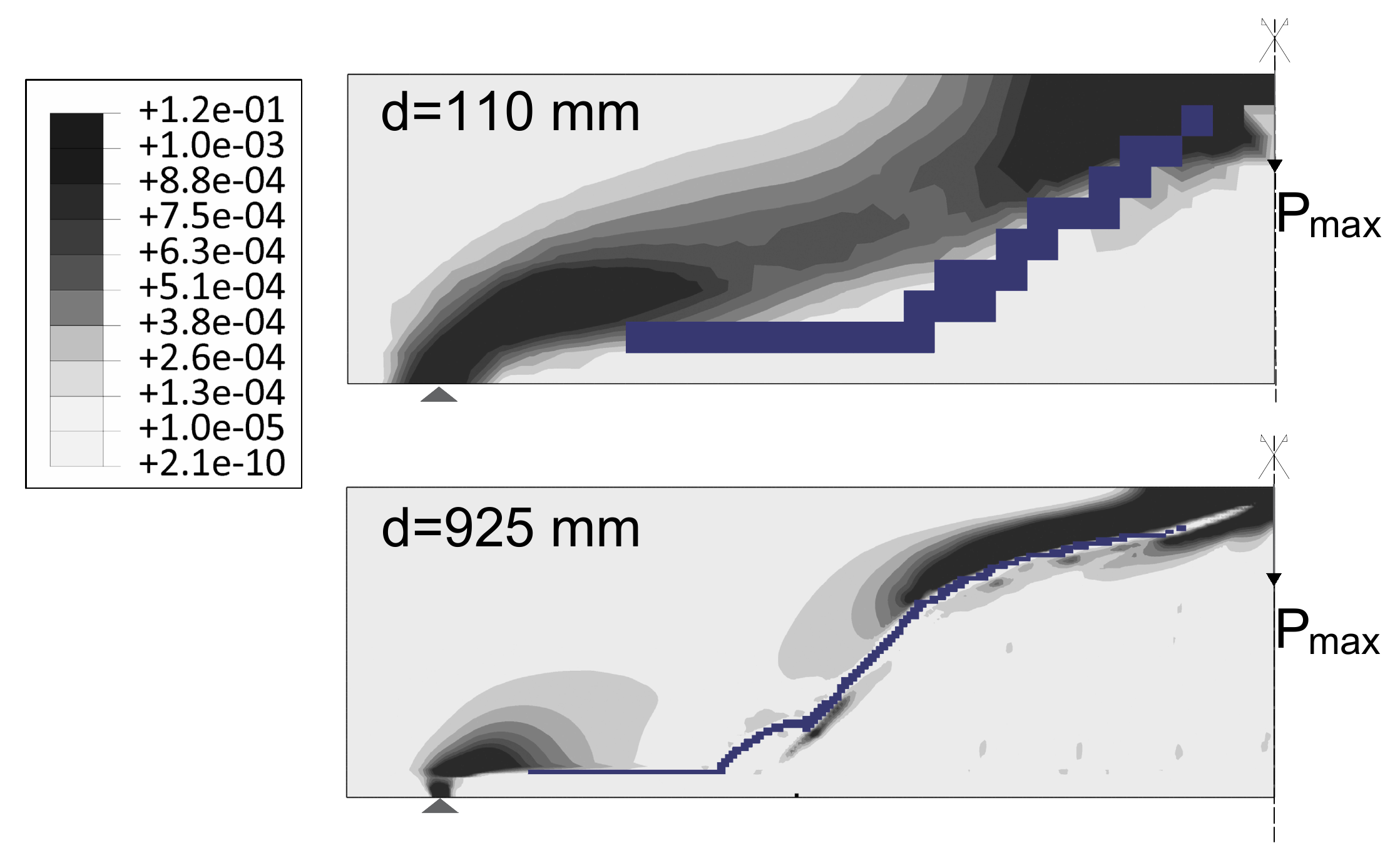}
\caption{\label{f3} \sf Strain energy density distribution at maximum load ($P_{max}$) calculated for Toronto test beams \cite{Toronto}.}
\efi

%4
\bfi
\centering
\includegraphics[width = 65mm, height=50mm]{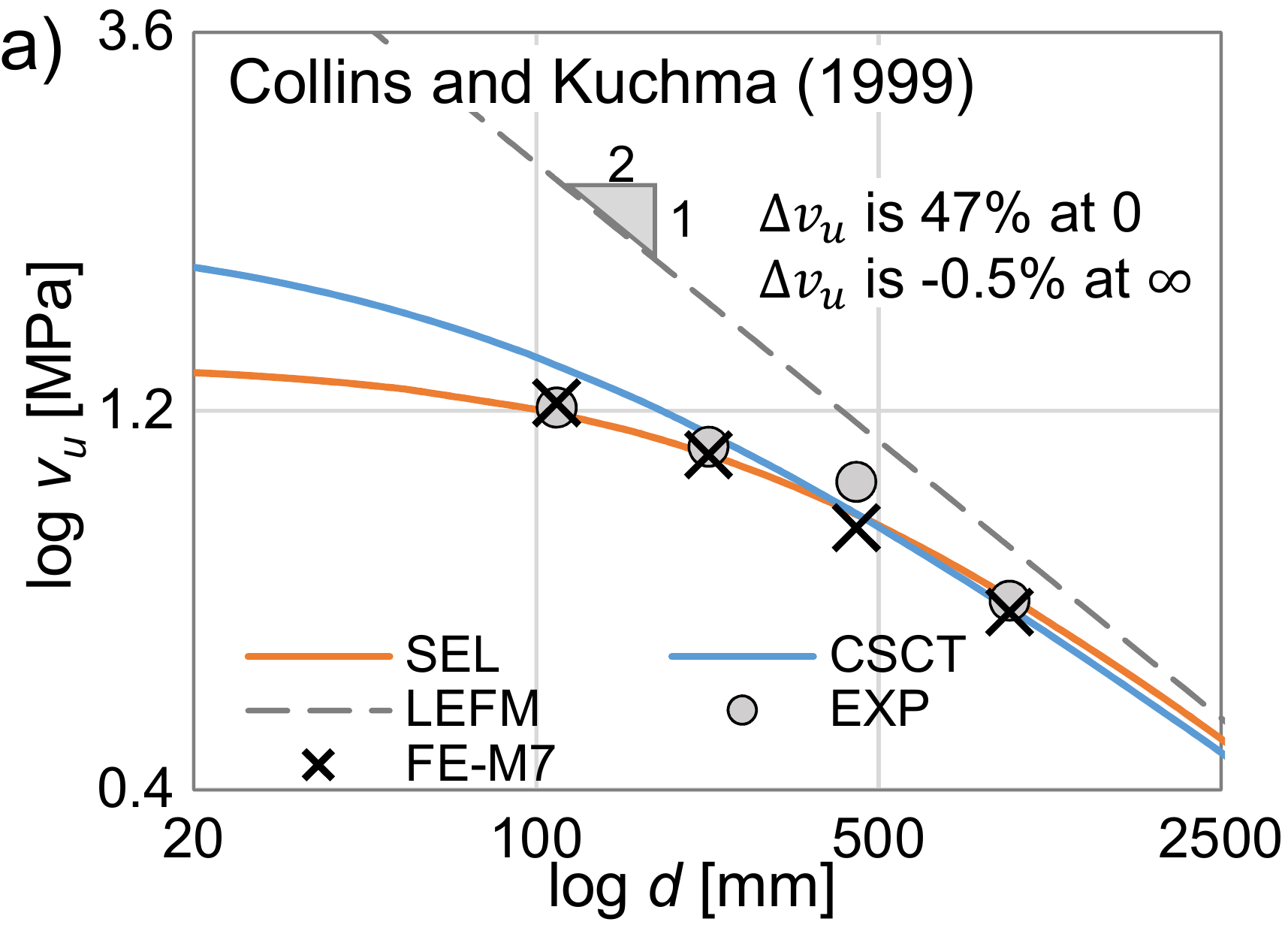}
\includegraphics[width = 65mm, height=50mm]{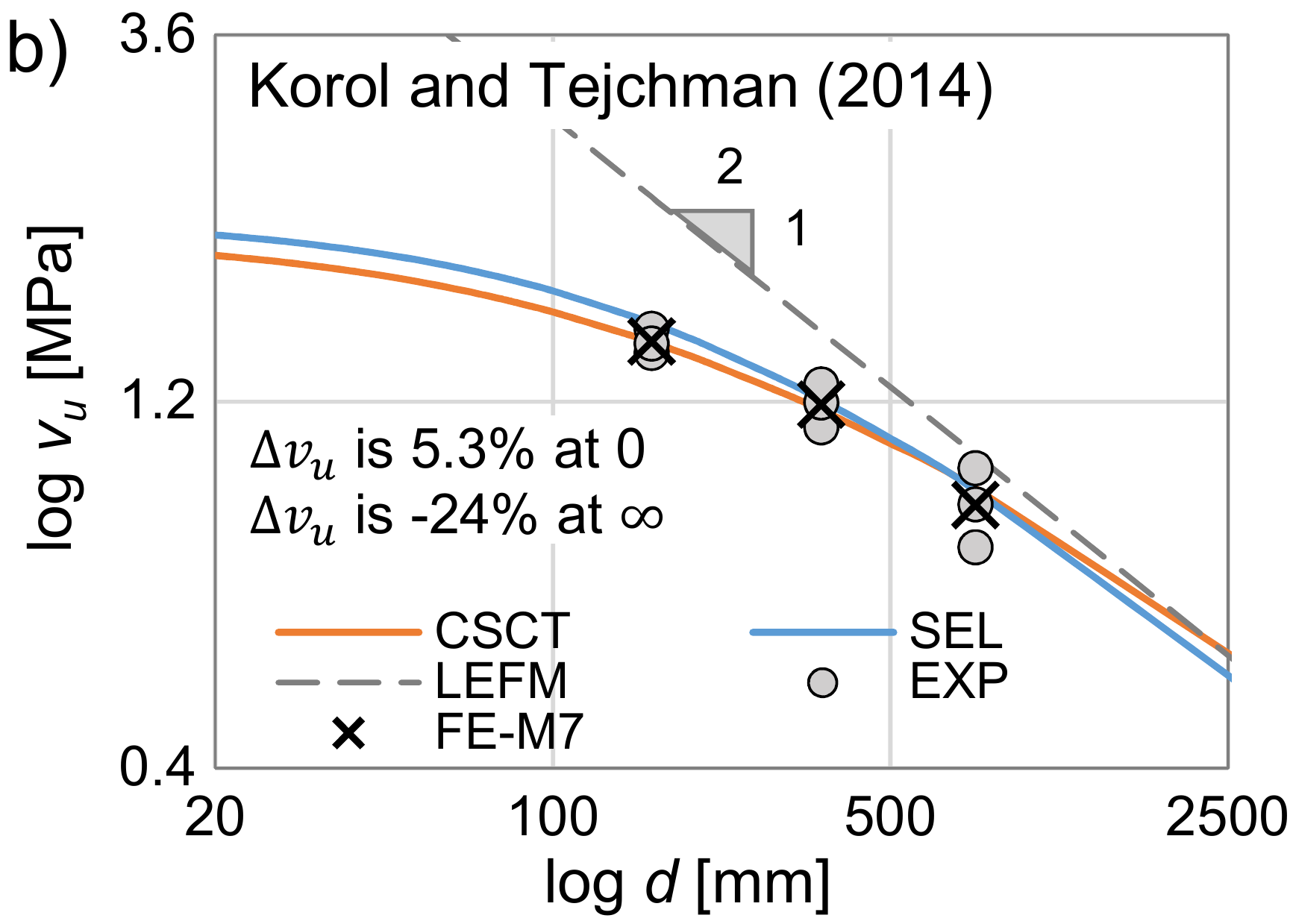}
\includegraphics[width = 65mm, height=50mm]{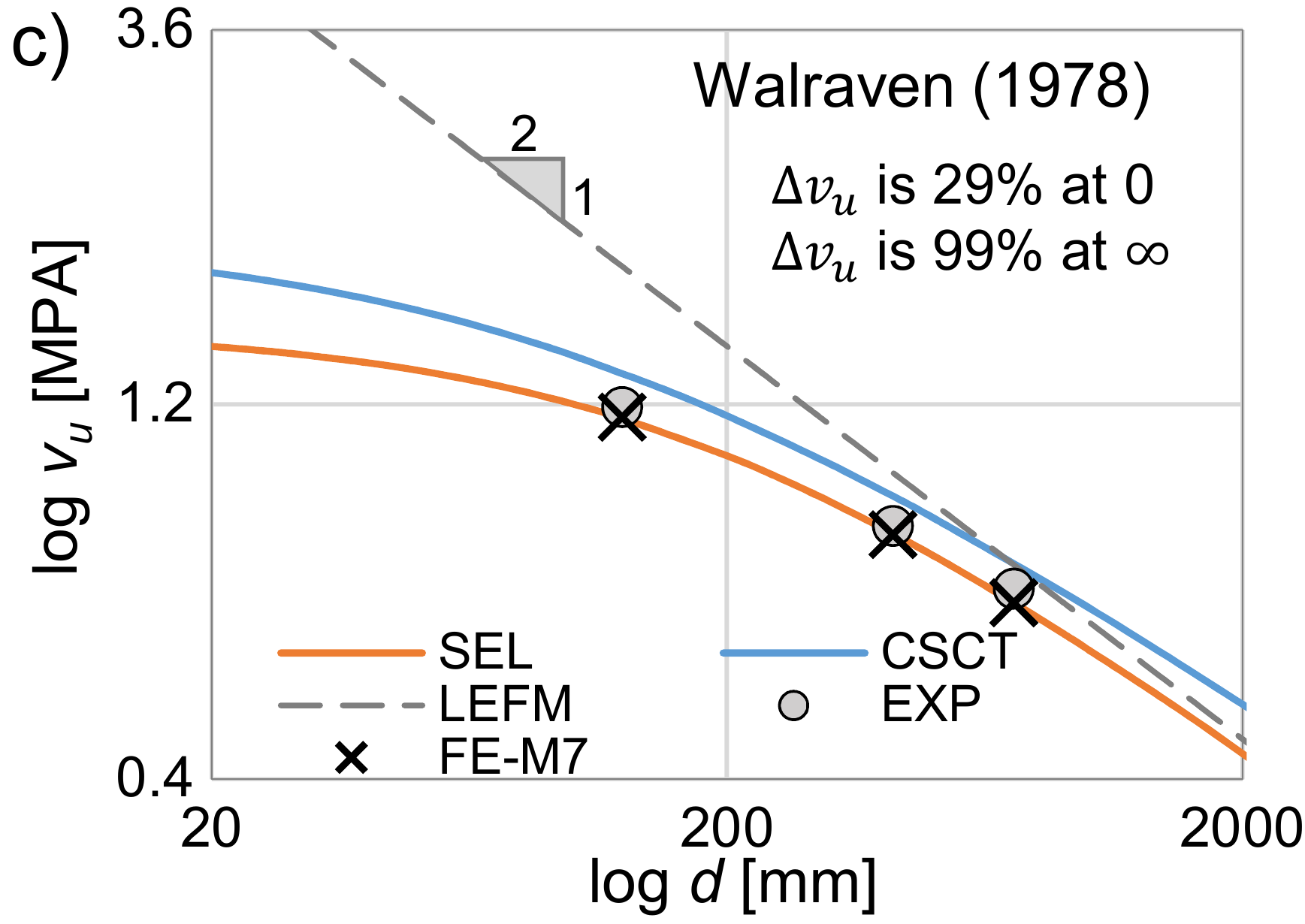}
\caption{\label{f4} \sf Comparison of test data: a) from \cite{Toronto}, b) from \cite{Korol14} and c) from \cite{Walraven} with FE results and with the size effect curves of SEL (Eq. \ref{8}) and CSCT (Eq. \ref{1}) ($\Del v_u$ are the percentage errors of CSCT compared to SEL fits)}
\efi

\end{document}